\documentclass[11pt]{article}
\usepackage{amsmath, amssymb, amsfonts, amsthm}
\usepackage{epsfig}

\newtheorem{theorem}{Theorem}[section]
\newtheorem{proposition}[theorem]{Proposition}

\begin{document}

\title{Derivation of effective evolution equations \\ from many body quantum dynamics}

\author{Benjamin Schlein\\
\\
DPMMS, University of Cambridge,\\
Cambridge, CB3 0WB, UK\\ 
E-mail: b.schlein@dpmms.cam.ac.uk}

\maketitle

\begin{abstract}
We review some recent results concerning the derivation of effective evolution equations from first principle quantum dynamics. In particular, we discuss the derivation of the Hartree equation for mean field systems and the derivation of the Gross-Pitaevskii equation for the time evolution of Bose-Einstein condensates.
\end{abstract}

\section{Introduction}

We consider a quantum mechanical system of $N$ spinless bosons in three dimensions. The system is described on the Hilbert space ${\cal H}_N = L^2_s ({\mathbb R}^{3N})$, consisting of all functions in $L^2 ({\mathbb R}^{3N})$ which are symmetric with respect to arbitrary permutations of the $N$ particles. The time evolution of a quantum mechanical system of $N$ bosons is governed by the $N$-particle Schr\"odinger equation \begin{equation}\label{eq:schr} i\partial_t \psi_{N,t} = H_N \psi_{N,t} \end{equation} for the wave function $\psi_{N,t} \in {\cal H}_N$ of the system. Here $H_N$ is a self-adjoint operator on ${\cal H}_N$, known as the Hamilton operator. We will consider Hamilton operators with two-body interactions described by a potential $V$, having the form \[ H_N = \sum_{j=1}^N  -\Delta_{x_j} + \sum_{i<j}^N V (x_i - x_j) \, . \] 

Typically, the value of $N$ in systems of physical interest is huge; it varies from $N \simeq 1000$ in very dilute experimental samples of Bose-Einstein condensates, up to $N \simeq 10^{23}$ in chemical experiments, or even to $N \simeq 10^{30}$ in astronomy. For such values of $N$, it is of course impossible to solve (\ref{eq:schr}) explicitly; numerical methods are completely useless as well. Fortunately, also from the point of view of physics, it is not so important to know the precise solution to (\ref{eq:schr}); it is much more important, for physicists performing experiments, to have information about the macroscopic properties of the system, which describe the typical behavior of the particles, and result from averaging over the $N$ particles. It turns out that the description of the macroscopic properties of the dynamics is particularly simple in so called mean field systems.

\section{Mean Field Systems}
\label{sec:MF}

A mean-field system is described by an $N$-body Hamilton operator of the form
\begin{equation}\label{eq:MFham}
H^{\text{mf}}_N = \sum_{j=1}^N -\Delta_j + \frac{1}{N} \sum_{i<j}^N V (x_i -x_j)
\end{equation}
acting on ${\cal H}_N$. The mean-field character of the Hamiltonian is expressed by the factor $1/N$ in front of the interaction; this factor guarantees that the kinetic and potential energies are both typically of order $N$ in the limit of large $N$.

We are going to study the solution $\psi_{N,t} = e^{-i H^{\text{mf}}_N t} \psi_N$ of the Schr\"odinger equation with factorized initial data $\psi_N = \varphi^{\otimes N}$, 
for some $\varphi \in L^2 ({\mathbb R}^{3})$.

In order to study the properties of $\psi_{N,t}$ in the limit $N \to \infty$, it is convenient to introduce marginal densities. For $k =1, \dots ,N$, the $k$-particle marginal density $\gamma_{N,t}^{(k)}$ associated with $\psi_{N,t}$ is defined by taking the partial trace of the density matrix $\gamma_{N,t} = |\psi_{N,t} \rangle \langle \psi_{N,t}|$ over the degrees of freedom of the last $N-k$ particles, 
\[ \gamma^{(k)}_{N,t} = \mbox{Tr}_{k+1, \dots , N} \gamma_{N,t} = \mbox{Tr}_{k+1, \dots ,N} |\psi_{N,t} \rangle \langle \psi_{N,t}| \, . \]
In other words, $\gamma^{(k)}_{N,t}$ is defined as the non-negative trace class operator on $L^2 ({\mathbb R}^{3 k})$ with kernel given by
\begin{equation}\label{eq:gammakNt}
\begin{split}
\gamma^{(k)}_{N,t} ({\bf x}_k ; {\bf x}'_k) &= \int {\mbox{d}} {\bf x}_{N-k} \, \overline{\psi}_{N,t} ({\bf x}_k , {\bf x}_{N-k}) \, \psi_{N,t} ({\bf x}'_k, {\bf x}_{N-k})\,.
\end{split}
\end{equation}

Due to the interaction, we cannot expect $\psi_{N,t}$ to be factorized. However, because of the mean-field character of the potential, it turns out that the $k$-particle marginal $\gamma^{(k)}_{N,t}$ associated with $\psi_{N,t}$ still factorizes in the limit $N \to \infty$, for arbitrary fixed $t \in {\mathbb R}$ and $k \in {\mathbb N}$. {F}rom the asymptotic factorization of the marginals, one can also obtain a self-consistent equation for the evolution of the condensate wave function (the one-particle orbital onto which the marginals factorize as $N\to\infty$), which turns out to be the nonlinear Hartree equation. 
\begin{theorem}\label{thm:MF} Under appropriate assumptions on the interaction potential $V$, let $\psi_N = \varphi^{\otimes N}$ for some $\varphi \in H^1 ({\mathbb R}^3)$ with $\| \varphi \|=1$, and $\psi_{N,t} = e^{-iH_N t} \psi_N$ (with $H_N$ defined in (\ref{eq:MFham})). Then, for every fixed $k \geq 1$ and $t \in {\mathbb R}$, we have
\begin{equation}\label{eq:conv} \gamma^{(k)}_{N,t} \to |\varphi_t \rangle \langle \varphi_t|^{\otimes k} \qquad \text{as } N \to \infty \end{equation} in the trace-norm topology. Here $\varphi_t$ is the solution to the nonlinear Hartree equation \begin{equation}\label{eq:hartree} i\partial_t \varphi_t = -\Delta \varphi_t + (V*|\varphi_t|^2 ) \varphi_t \end{equation} with initial data $\varphi_{t=0} = \varphi$.
\end{theorem}

The first proof of this theorem was obtained by Spohn in \cite{Sp} under the assumption of a bounded potential. The approach introduced by Spohn is based on the study of the time evolution of the reduced density matrices, which is governed by the BBGKY Hierarchy (for $k = 1, \dots, N$)
\begin{equation}\label{eq:BBGKY}
\begin{split}
i \partial_t \gamma^{(k)}_{N,t}  = \; &\sum_{j=1}^k \left[-\Delta_{x_j} , \gamma^{(k)}_{N,t} \right] + \frac{1}{N} \sum_{i < j}^k \left[ V (x_i - x_j), \gamma^{(k)}_{N,t} \right] \\ &+ \left(1- \frac{k}{N} \right) \sum_{j=1}^k \mbox{Tr}_{k+1} \, \left[ V(x_j - x_{k+1}) , \gamma^{(k+1)}_{N,t} \right]
\end{split}
\end{equation}
where $\mbox{Tr}_{k+1}$ denotes the partial trace over the variable $x_{k+1}$, and where we used the convention $\gamma^{(N+1)}_{N,t} \equiv 0$. If we fix $k \geq 1$ and we let $N \to \infty$, the BBGKY Hierarchy formally converges to the infinite hierarchy
\begin{equation}\label{eq:infhier}
i \partial_t \gamma^{(k)}_{\infty,t}  = \sum_{j=1}^k \left[-\Delta_{x_j} , \gamma^{(k)}_{\infty,t} \right] + \sum_{j=1}^k \mbox{Tr}_{k+1} \, \left[ V(x_j - x_{k+1}) , \gamma^{(k+1)}_{\infty,t} \right] \, .
\end{equation}
It is then worth noticing that this infinite hierarchy has a factorized solution. In fact, the family $\gamma^{(k)}_{\infty,t} = |\varphi_t \rangle \langle \varphi_t|^{\otimes k}$ solves (\ref{eq:infhier}) if and only if $\varphi_t$ solves the nonlinear Hartree equation (\ref{eq:hartree}). This simple observation implies that in order to obtain a rigorous proof of Theorem \ref{thm:MF}, it is enough to complete the following three steps.
\begin{itemize}
\item[$\bullet$] Prove the {\it compactness} of the sequence of densities $\gamma^{(k)}_{N,t}$ with respect to an appropriate weak topology.
\item[$\bullet$] Prove the {\it convergence} to the infinite hierarchy. In other words, show that every limit point of the sequence $\gamma^{(k)}_{N,t}$ solves (\ref{eq:infhier}).
\item[$\bullet$] Prove the {\it uniqueness} of the solution of the infinite hierarchy. 
\end{itemize}
These three steps immediately imply that $\gamma^{(k)}_{N,t} \to |\varphi_t \rangle \langle \varphi_t|^{\otimes k}$ as $N \to \infty$ (the convergence is first in a weak sense, but since the limit is a rank one projection, this automatically implies convergence in the trace norm topology).

Implementing this three step strategy becomes more difficult for potentials with singularities. In \cite{EY}, Erd\H os and Yau extended Spohn's result to the case of a Coulomb potential $V(x) = \pm 1/|x|$. To deal with the singularity of $V$, they proved that arbitrary limit points $\gamma^{(k)}_{\infty,t}$ of the sequence of reduced densities $\gamma_{N,t}^{(k)}$ satisfy strong a-priori bounds of the form \begin{equation}\label{eq:aprik} \mbox{Tr} \, (1-\Delta_{x_1}) \dots (1-\Delta_{x_k}) \gamma^{(k)}_{\infty,t} \leq C^k \end{equation} for some finite constant $C>0$ and every $k \in {\mathbb N}, t\in {\mathbb R}$. Hence, it is enough to prove the uniqueness of the solution of (\ref{eq:infhier}) in the class of densities satisfying (\ref{eq:aprik}), where the Coulomb singularity can be controlled by the operator inequality $|x|^{-1} \leq C (1-\Delta)$. A similar technique was used in \cite{ES}, a joint work with A. Elgart, to handle bosons with a relativistic dispersion law interacting through a mean field Coulomb potential (the relativistic case is more complicated, because one can only prove weaker a-priori estimates).

A disadvantage of the strategy exposed above is the lack of effective bounds on the rate of convergence in (\ref{eq:conv}). Following a different approach, proposed initially by Hepp in the slightly different context of the classical limit of quantum mechanics and later extended by Ginibre and Velo to a larger class of potentials (see \cite{He, GV}), we recently managed in \cite{RS}, a joint work with I. Rodnianski, to obtain effective bounds on the difference between the full Schr\"odinger evolution and the Hartree approximation.
For potentials with at most a Coulomb singularity, we show that there exist constants $C,K>0$ such that 
\begin{equation}\label{eq:bdtr} \mbox{Tr} \, \left| \gamma^{(1)}_{N,t} - |\varphi_t \rangle \langle \varphi_t| \right|  \leq \frac{C e^{Kt}}{\sqrt{N}} \end{equation}
for all $t \in {\mathbb R}$ (similar bounds holds for higher order marginals as well). Note that the bound on the r.h.s. of (\ref{eq:bdtr}) is not expected to be optimal in its $N$ dependence (in fact, for bounded potential, it was proven in \cite{ErS}, a joint work with L. Erd\H os, that the l.h.s. is at most of the order $1/N$ for every fixed $t$). The proof of (\ref{eq:bdtr}) is based on a Fock space representation of the many boson system, and on the study of the time evolution of coherent states.

Recently, further progress has been ahieved in the analysis of the dynamics of mean-field systems. In \cite{KP}, Knowles and Pickl obtain effective estimates on the rate of convergence to the Hartree equation, for potential with strong singularity (their proof is based on a method developed by Pickl in \cite{P}). In \cite{GMM}, Grillakis, Machedon, and Margetis found second order correction to the mean-field evolution of coherent states.

\section{Dynamics of Bose-Einstein condensates}

Another class of systems for which an effective evolution equation can be derived consists of dilute Bose gases in the so called Gross-Pitaevskii scaling limit. The Hamiltonian of a trapped dilute Bose gas is given by \begin{equation}\label{eq:trapGPham} H_N^{\text{trap}} = \sum_{j=1}^N \left(-\Delta_{x_j} + V_{\text{ext}} (x_j)  \right)+ \sum_{i<j}^N V_N (x_i -x_j)
\end{equation}
with $V_N (x) = N^2 V (Nx)$. Here $V_{\text{ext}}$ is a confining potential, while $V \geq 0$ is a short range potential describing the interaction among the particles. We will denote by $a_0$ the scattering length of $V$. Recall that the scattering length is defined (for integrable potentials) as
\begin{equation}\label{eq:scatt1}
8 \pi a_0 = \int V(x) f(x) {\mbox{d}} x 
\end{equation}
where $f(x)$ is the solution of the zero-energy scattering equation
\begin{equation}\label{eq:0en} \left(-\Delta + \frac{1}{2} V \right) f = 0 \end{equation} with the boundary condition $f(x) \to 1$, as $|x| \to \infty$. Note that, if $a_0$ is the scattering length of $V$, the scattering length of the rescaled potential $V_N$ is exactly given by $a = a_0/N$. This follows by simple scaling because, if $f$ is the solution of (\ref{eq:0en}), then $f_N (x) = f (Nx)$ solves the rescaled problem \begin{equation}\label{eq:0enN}  \left(-\Delta + \frac{1}{2} V_N \right) f_N = 0 \, . \end{equation}

In \cite{LSY}, Lieb, Seiringer, and Yngvason proved that, if $E_N$ denotes the ground state energy of $H_N^{\text{trap}}$,
\[ \lim_{N \to \infty} \frac{E_N}{N} = \min_{\varphi \in L^2 ({\mathbb R}^3) : \| \varphi \|= 1} \, {\cal E}_{\text{GP}} (\varphi) \]
where ${\cal E}_{\text{GP}} (\varphi)$ is the so called Gross-Pitaevskii energy functional given by
\begin{equation}\label{eq:GPfunc} {\cal E}_{\text{GP}} (\varphi) =  \int {\mbox{d}} x \,  \left( |\nabla \varphi (x)|^2 + |V_{\text{ext}} (x)|^2 + 4 \pi a_0 |\varphi (x)|^4 \right)  \, .\end{equation}
In \cite{LS}, Lieb and Seiringer also proved that the ground state $\psi_N^{\text{trap}}$ of the Hamiltonian $H_N^{\text{trap}}$ exhibits complete Bose-Einstein condensation. More precisely, they proved that, if $\gamma^{(1)}_N$ denotes the one-particle marginal associated with $\psi_N^{\text{trap}}$, \[ \lim_{N \to \infty} \gamma^{(1)}_N = |\phi_{\text{GP}} \rangle \langle \phi_{\text{GP}} | \] where $\phi_{\text{GP}}$ denotes the minimizer of (\ref{eq:GPfunc}) (the limit is in the trace norm topology). The interpretation of this result is straightforward; in the ground state of $H_N^{\text{trap}}$ all particles, up to a fraction which vanishes in the limit $N \to \infty$, are in the same one-particle state with orbital $\phi_{\text{GP}}$.

The question now is what happens to the condensate when the trapping potential is turned off. The next theorem, proven in a series of joint works with L. Erd\"os and H.-T. Yau (see \cite{ESY1,ESY2,ESY3,ESY4}) shows that complete condensation is preserved by the Schr\"odinger dynamics and that the Gross-Pitaevskii theory correctly predicts the  time evolution of the condensate wave function.

\begin{theorem}\label{thm:GP}
Suppose $V \geq 0$ is bounded and decays sufficiently fast and let
\begin{equation}\label{eq:GPham} H_N = \sum_{j=1}^N -\Delta_{x_j} + \sum_{i<j}^N V_N (x_i - x_j) \qquad \text{with } V_N (x) = N^2 V(Nx) \, . \end{equation} Consider a family of initial data $\psi_N$ with finite energy per particle \[ \langle \psi_N , H_N \psi_N \rangle \leq C N \] and such that the one-particle marginal $\gamma^{(1)}_N$ associated with $\psi_N$ satisfy \[ \gamma^{(1)}_N \to |\varphi \rangle \langle \varphi| \qquad \text{as } N \to \infty \] for some $\varphi \in H^1 ({\mathbb R}^3)$ with $\| \varphi \|=1$. Let $\psi_{N,t} = e^{-iH_N t} \psi_N$ and $\gamma^{(k)}_{N,t}$ be the $k$-particle marginal associated with $\psi_{N,t}$. Then, for every fixed $t \in {\mathbb R}$, and every $k \geq 1$, we have \[ \gamma_{N,t}^{(k)} \to |\varphi_t \rangle \langle \varphi_t|^{\otimes k} \qquad \text{as } N \to \infty \] in the trace norm topology. Here $\varphi_t$ is the solution to the time dependent Gross-Pitaevskii equation \begin{equation}\label{eq:GP} 
i\partial_t \varphi_t = -\Delta \varphi_t + 8 \pi a_0 |\varphi_t|^2 \varphi_t \end{equation} with initial data $\varphi_{t=0} = \varphi$.
\end{theorem}

Observe that the Hamiltonian (\ref{eq:GPham}) can be rewritten as \[ H_N = \sum_{j=1}^N -\Delta_{x_j} + \frac{1}{N} \sum_{i<j}^N v_N (x_i - x_j) \] with $v_N (x) = N^3 V (Nx)$.  Formally, we have $v_N (x) \to b_0 \delta (x)$ with $b_0 = \int V(x) dx$ as $N \to\infty$. It is therefore tempting to interpret (\ref{eq:GPham}) as a mean field Hamiltonian with a potential which depends on $N$ and tends, as $N \to \infty$, to a delta-function. However, one should be very careful with this interpretation; physically, mean field systems are characterized by very frequent and very weak interactions. The Hamiltonian (\ref{eq:GPham}), on the other hand, describes a completely different regime, with very rare collisions (particles only interact when they are at distances of order $N^{-1}$,  much smaller than the typical interparticle distance) and very strong.

To understand the origin of the Gross-Pitaevskii equation (\ref{eq:GP}) as the effective equation for the dynamics of the condensate wave function, consider the equation \begin{equation}\label{eq:BBG1} i\partial_t \gamma^{(1)}_{N,t} = \left[ -\Delta , \gamma^{(1)}_{N,t} \right] + (N-1) \mbox{Tr}_2 \left[ V_N (x_1 -x_2) , \gamma^{(2)}_{N,t} \right] \end{equation} which describes the evolution of the one-particle marginal $\gamma^{(1)}_{N,t}$ associated with the solution $\psi_{N,t}$ of the $N$ particle Schr\"odinger equation (this is the first equation of the BBGKY hierarchy for the family $\{ \gamma^{(k)}_{N,t} \}_{k=1}^N$). If condensation is preserved by the time evolution, we should expect that, for large $N$, $\gamma^{(k)}_{N,t} \simeq |\varphi_t \rangle \langle \varphi_t|^{\otimes k}$ for both $k=1,2$. Inserting this ansatz in (\ref{eq:BBG1}), and using that $(N-1)V_N (x) \simeq N^3 V (N x) \to b_0 \delta (x)$ as $N \to \infty$, we obtain the self-consistent equation \[ i\partial_t \varphi_t = -\Delta \varphi_t + b_0 |\varphi_t|^2 \varphi_t \] for $\varphi_t$. This equation has the same form as (\ref{eq:GP}), but a different constant in front of the nonlinearity. We get the wrong constant because, in this naive argument, we neglected the correlations characterizing the two-particle density $\gamma^{(2)}_{N,t}$. It turns out that the solution of the Schr\"odinger equation develops a short scale correlation structure which varies exactly on the same length scale of order $N^{-1}$ characterizing the interaction potential. If we assume for a moment that the correlations can be described by the solution of the zero-energy scattering equation $f_N$ (see \ref{eq:0enN}), we may expect that for large but finite $N$ the kernel of the one- and the two-particle marginals can be approximated by \begin{equation}\label{eq:ans2} \begin{split} \gamma^{(1)}_{N,t} (x_1; x'_1) &\simeq \varphi_t (x_1) \overline{\varphi}_t (x'_1) \\  \gamma^{(2)}_{N,t} (x_1, x_2 ; x'_1, x'_2) &\simeq f_N (x_1 -x_2) f_N (x'_1 - x'_2) \varphi_t (x_1) \varphi_t (x_2) \overline{\varphi}_t (x'_1) \overline{\varphi}_t (x'_2) \, .\end{split}\end{equation} Inserting this new, more precise, ansatz in (\ref{eq:BBG1}), we obtain another self-consistent equation for $\varphi_t$, which, by (\ref{eq:scatt1}), is exactly the Gross-Pitaevskii equation (\ref{eq:GP}). 

{F}rom this heuristic discussion it is already clear that in order to obtain a rigorous proof of Theorem \ref{thm:GP}, one of the main steps consists in showing that the solution of the $N$-particle Schr\"odinger equation develops a short scale correlations structure, and that this structure can be described, in very good approximation, by the solution of the zero energy scattering equation $f_N$. In order to reach this goal, we make use of strong a-priori bounds of the form \begin{equation}\label{eq:apriori} \int {\mbox{d}} {\bf x} \left| \nabla_{x_i} \nabla_{x_j} \frac{\psi_{N,t} ({\bf x})}{f_N (x_i -x_j)}\right|^2 \leq C \end{equation} valid for all $i \not = j$, uniformly in $t$ and in $N$, and for every initial wave function $\psi_N$ such that $\langle \psi_N, H_N^2 \psi_N \rangle \leq CN^2$ (the bounds (\ref{eq:apriori}) were derived in \cite{ESY2} under the assumption of sufficiently weak interaction potential; the case of strong potential was treated in \cite{ESY3}, using different a-priori bounds). The estimate (\ref{eq:apriori}), which is proven using the conservation of the expectation of the Hamiltonian squared, identifies the short scale correlation structure of $\psi_{N,t}$. This can be used to prove the convergence of solutions of the BBGKY hierarchy towards an infinite hierarchy of equation similar to (\ref{eq:infhier}), but with $V$ replaced by $8\pi a_0$. 

To complete the proof of Theorem \ref{thm:GP} (following the strategy outlined in Section~\ref{sec:MF}), one still needs to prove the uniqueness of the solution of the infinite hierarchy; technically, this is actually the most difficult part of the proof. On the one hand, we have to prove that the limit point of the densities satisfy a-priori estimates of the form (\ref{eq:aprik}); the problem here is much more involved compared with the case of a mean-field Coulomb interaction because of the presence of the singular correlation structure. On the other hand, when we prove the uniqueness of the solution of the infinite hierarchy in the class of densities satisfying the a-priori estimates, we have to face the problem that, in three dimensions, the delta-interaction is not bounded by the Laplacian (in contrast to $|x|^{-1} \leq C(1-\Delta)$, the inequality $\delta (x) \leq C (1-\Delta)$ is not true in three dimensions). Details of the proof of the a-priori bounds can be found in \cite[Section 5]{ESY2}. The proof of the uniqueness (in the class of densities satisfying the a-priori bounds) can be found in \cite{ESY1}. 

\section{Dynamical Formation of Correlations}

In the last section, we stressed the fact that the emergence of the scattering length in the Gross-Pitaevskii equation (\ref{eq:GP}) is a consequence of the short scale correlation structure characterizing the solution $\psi_{N,t}$ of the $N$-particle Schr\"odinger equation. If one assumes that $\langle \psi_N , H_N^2 \psi_N \rangle \leq C N^2$, the presence of the correlation structure in the evolved wave function $\psi_{N,t}$ follows from a-priori bounds of the form (\ref{eq:apriori}).
It turns out, however, that Theorem \ref{thm:GP} can also be applied to completely factorized initial data of the form $\psi_N = \varphi^{\otimes N}$. Therefore, also for initial data with absolutely no correlations among the particles, the evolution of the condensate wave function is described by the Gross-Pitaevskii equation (\ref{eq:GP}) with coupling constant proportional to $a_0$. This observation suggests that the time evolution $\psi_{N,t}$ of the completely factorized data $\psi_N$ develops the short scale correlation structure within very small time intervals, whose length vanishes in the limit $N \to \infty$. Note that the a-priori bound (\ref{eq:apriori}) cannot be used to prove that 
$\psi_{N,t}$ contains the correct short scale structure because $\langle \varphi^{\otimes N}, H_N^2 \varphi^{\otimes N} \rangle \simeq N^3 \gg N^2$. In order to prove Theorem \ref{thm:GP} for factorized initial data $\psi_N$ we need therefore an approximation argument to replace $\psi_N$ by the $N$-particle wave function $\widetilde{\psi}_{N, \varepsilon} \simeq \chi (H_N \leq N \varepsilon^{-1}) \psi_N$ having the correct correlation structure. We perform then the whole analysis on the evolution of $\widetilde{\psi}_{N,\varepsilon}$ and only at the end, after taking $N \to \infty$, we let $\varepsilon \to 0$.

In particular, our analysis does not give any proof of the fact that the evolution of a factorized initial data develops a short scale correlation structure within short times. 
In this section, which is based on \cite{EMS}, a joint work with L. Erd\H os and A. Michelangeli, we provide some direct evidence supporting this claim.

It turns out that the dynamical formation of correlation can be understood as a two-body phenomenon. Consider the two-particle Hamiltonian \[{\frak h}_N^{(1,2)} = -\Delta_1 - \Delta_2 + N^2 V (N (x_1 - x_2)) \] and the factorized two-particle wave function $\psi (x_1, x_2) = \varphi (x_1) \varphi (x_2)$ for a $\varphi \in L^2 ({\mathbb R}^3)$ smooth and decaying fast enough. In order to monitor the formation of correlations in a window of length $N^{-1} \leq \ell \ll 1$, we introduce the quantity \[ {\cal F}_N (t) = \int {\mbox{d}} x_1 {\mbox{d}} x_2 \, \theta (|x_1 - x_2| \leq \ell) \, \left| \nabla_{x_1 -x_2} \frac{ e^{-it {\frak h}_N^{(1,2)}} \psi (x_1, x_2)}{f_N (x_1 -x_2)} \right|^2 \, . \] 
At time $t=0$, $\psi$ is factorized and it is simple to show that $N {\cal F}_N (0) \simeq 1$. The following proposition gives an upper bound on ${\cal F}_N (t)$ for $t >0$. 
\begin{proposition}\label{prop:two} Suppose that $0 \leq t \leq N^{-1}$ and $N^{-1} \leq \ell \ll 1$.
Then 
\[ N {\cal F}_N (t) \leq \frac{(\log N^2 t)^6}{N^2 t} \, (N \ell)^3 \, . \] 
\end{proposition}
In particular, Proposition \ref{prop:two} implies that, if $\ell \simeq N^{-1}$, ${\cal F}_N (t) \ll {\cal F}_N (0)$ for all $N^{-2} \ll t \leq N^{-1}$; this is clear evidence for the formation of the correlation structure on length scale of order $\ell \simeq N^{-1}$ within times of order $N^{-2}$ (note that, for technical reasons, the result is restricted to times $t \leq N^{-1}$; hence, strictly speaking, Proposition \ref{prop:two} only shows that correlations form during the time interval $N^{-2} \ll t \leq N^{-1}$).

To prove Proposition \ref{prop:two} it is useful to switch to a center of mass coordinate $\eta = (x_1 + x_2)/2$ and to a relative coordinate $x= x_2 - x_1$. In these coordinates, the Hamiltonian takes the form \[ {\frak h}_N^{(1,2)} = -\frac{\Delta_\eta}{2} - 2 \Delta_x + V_N (x) =: -\frac{\Delta_\eta}{2} + {\frak h}_N \]
with ${\frak h}_N = -2\Delta_x + V_N (x)$. If we moreover rescale the relative coordinate letting $X = Nx$, we find \[ {\frak h}_N = N^2 (-2\Delta_X + V(X)) =: N^2 {\frak h} \, .\] Hence, setting $T = N^2 t$, we find \begin{equation} \label{eq:Xeta} N {\cal F}_N (t) = \int {\mbox{d}} \eta {\mbox{d}} X \, \theta (|X| \leq N \ell) \,  \left| \nabla_X \frac{(e^{-iT {\frak h}} \psi_N ) (\eta, X) }{f(X)} \right|^2 \end{equation} where $\psi_{N} (\eta, X) = \varphi (\eta + X/N) \varphi (\eta -X/N)$. In \eqref{eq:Xeta}, the $N$-dependence is hidden in the initial data $\psi_N$ which has the property of being essentially constant in $X$, as long as $|X| \ll N$. The formation of correlations for $T \gg 1$ is then a statement about the local relaxation of an initial data which is essentially independent of $X$, under the dynamics generated by ${\frak h}$, to the function $f (X)$. Recall here that $f$ is such that ${\frak h} f =0$; one can prove that, for large $|X|$, it has the form $f(X) \simeq 1 - a_0/|X|$. Defining $\omega (X) = 1 - f(X)$, we have, very formally,
\[ e^{-iT{\frak h}} 1 = e^{-iT{\frak h}} (1-\omega) + e^{-iT{\frak h}} \omega = f + e^{-iT{\frak h}} \omega \, . \]
The second term disperses away for large $T$ and therefore, if we only look inside a window of order one ($\ell \simeq N^{-1}$), $e^{-iT{\frak h}} 1 \simeq f$. This explains why the quantity under investigation becomes small for $T \gg 1$ (which is equivalent to $t \gg N^{-2}$). The main obstacle to make this formal argument rigorous is the lack of decay of $\omega$. Since $\omega \simeq |X|^{-1}$ at infinity, the standard dispersion estimates for the evolution of $\omega$ cannot be applied. Instead, we show new dispersion estimates of the form 
\begin{equation}\label{eq:disp} \| e^{-i\Delta t} \varphi \|_{\infty} \leq C t^{-\frac{3}{2s}} \left( \| \varphi \|_s + \| \nabla \varphi \|_{\frac{3s}{s+3}} + \| \nabla^2 \varphi \|_{\frac{3s}{2s+3}} \right) \end{equation}
for all $s \geq 3/2$ which require less decay of the initial data; the price to pay is that one needs some regularity (similar bounds can be obtained for the $L^q$-norms of $e^{-i\Delta t} \varphi$ for sufficiently large $q$). Choosing $s \geq 3$, the bound (\ref{eq:disp}) can be applied to the function $\omega$. Using Yajima bounds on the wave operator (see \cite{Ya}), estimates of the form (\ref{eq:disp}) can then be proven to hold also for the dynamics generated by ${\frak h}$.

Proposition \ref{prop:two} proves that the formation of correlation is a two-particle phenomenon. In order to understand the formation of correlation in the many particle setting, we would need to control the effect of the three-body interactions on the correlation structure. This is a very difficult task and, in general, not much can be rigorously established about this problem. Since, however, the many body system is very dilute, three body interactions are very rare. Therefore we can find an interval of time during which two body collisions are already effective (and lead therefore to the formation of the correlation structure) while three body collisions are still negligible. 
\begin{theorem}
Let \[ {\cal G}_N (t) = \int {\mbox{d}} {\bf x} \, \theta (|x_1 -x_2| \leq \ell) \, \left| \frac{\psi_{N,t} ({\bf x})}{f_N (x_1 -x_2)} - \psi_N ({\bf x}) \right|^2 \]
for $\psi_N = \varphi^{\otimes N}$ with $\varphi \in L^2 ({\mathbb R}^3)$ smooth and decaying fast enough, and $\psi_{N,t} = e^{-iH_N t} \psi_N$, with $H_N$ defined in (\ref{eq:GPham}). Then, for all $0 \leq t \leq N^{-1}$, \[ {\cal G}_N (t) \leq C {\cal G}_N (0) \bigg( \frac{(\log N)^{\frac{4}{5}}}{N^{\frac{1}{5}}}\frac{(N^2t)^2}{N\ell}+ \, \frac{\:(N\ell)^4}{N^2t} \left(\log N^2t \right)^6 \bigg) \, .\] 
\end{theorem}
In particular, the theorem implies that, when considering windows of size $\ell \simeq N^{-1}$, 
\[ {\cal G}_N (t) \ll \, {\cal G}_N(0) \qquad\textrm{if}\qquad 1 \; \ll \; N^2 t \; \ll \; N^{\frac{1}{10}} \,. \] This inequality indicates that, under the many body dynamics, a completely factorized initial data develops the short scale correlation structure within times of order $N^{-2}$ and keeps it at least up to times of order $N^{-1.9}$ (of course, we expect the correlation structure to remain intact up to times of order one, but proving this claim would require a better understanding of the effect of the many-body interactions).

\end{document}